\documentstyle[aps,12pt]{revtex}

\begin{document}

\title{Interdimensional degeneracies for\\
a quantum three-body system in $D$ dimensions}

\author{Xiao-Yan Gu \thanks{Electronic address:
guxy@mail.ihep.ac.cn} and Zhong-Qi Ma \thanks{Electronic address:
mazq@sun.ihep.ac.cn}}

\address{CCAST (World Laboratory), P.O.Box 8730, Beijing 100080, China \\
and Institute of High Energy Physics, Beijing 100039, China}

\author{Bin Duan \thanks{Electronic address: duanb@mpipks-dresden.mpg.de}}

\address{Max Planck Institute for the Physics of Complex Systems, \\
Noethnitzer Str. 38, 01187, Dresden, Germany}


\maketitle

\vspace{5mm}

\begin{abstract}

A new approach is developed to derive the complete spectrum of
exact interdimensional degeneracies for a quantum three-body
system in $D$-dimensions. The new method gives a generalization 
of previous methods.

\end{abstract}

\section{Introduction}

The property of quantum systems in high dimensions and its
relationship with that in three dimensions has aroused great
interest in statistic mechanics, particle physics, and nuclear
physics \cite{wil,rud,loe,yaf,her}. A characteristic feature for a
quantum few body system in $D$ dimensions is existence of exact
interdimensional degeneracies. For any central force problem in D
dimensions an isomorphism exists between angular momentum $L$ and
dimension $D$ such that each unit increment in $L$ is equivalent
to two-unit increment in D, as perhaps first noticed by Van Vleck
\cite{van}. For one-electron system states related by the
dimensional link $D,~l \leftrightarrow (D-2),~(l+1)$ are exactly
degenerate \cite{her2}. For two-electron system Herrick and
Stillinger found exact interdimensional degeneracies between the
states $^{1, 3}P^e$ and $^{1, 3}D^o$ in $D=3$ and the states $^{3,
1}S^e$ and $^{3, 1}P^o$ in $D=5$, respectively \cite{her3}. These
interdimensional degeneracies
\cite{her,her2,her3,goo,goo2,dor,dun1,ros,tan} were supposed
\cite{dun1,goo} to permit that the accurate energies for the
states $^{1, 3}P^e$ and $^{1, 3}D^o$ in $D=3$ may be calculated
with reduced effort by implementing the same procedures which
calculate the states $^{3, 1}S^e$ and $^{3, 1}P^o$ in $D=5$,
respectively. The pattern of approximate interdimensional
degeneracies among doubly excited states of two-electron atoms has
also been nicely elucidated \cite{ros} by means of the molecular
orbital description. This link, between dimension and orbital
angular momentum as a pervasive feature, can be used to classify
groups of quasidegenerate doubly excited atomic energies and to
explain striking similarities among certain pairs of
hyperspherical or molecular-orbital two-electron potential curves
\cite{ros}. Interdimensional degeneracies can also be used to
construct part of the D-dimensional spiked harmonic oscillator and
anharmonic oscillator bound-state spectra \cite{oma,oma2}.

As is well known, dimensional scaling methods provide a powerful
approach for studying atomic and molecular systems in D dimensions
\cite{her,cha,ger,pop,loe2,kai,kai2}, but much work has been
confined to S-wave states and some $P^e$ states
\cite{her2,her3,goo} which were obtained by exploiting a known
interdimensional degeneracies between $S^e$ states in 5 dimensions
and $P^e$ states in 3 dimensions. Applications of dimensional
scaling methods to higher angular momentum states of
multi-electron systems are sparse
\cite{her,her2,her3,goo,goo2,dor,dun1,ros,tan}. Recently, by the
method of group theory, Dunn and Watson developed a formalism for
the N-electron D-dimensional Schwartz expansion \cite{dun2,dun3},
and applied the principle method to the Schr\"{o}dinger equation
for two-electron system in an arbitrary $D$-dimensional space
\cite{dun4,dun5}. The resulting set of coupled differential
equations in the internal variables, enabling the methods of
dimensional scaling to be applied to higher-angular-momentum
states. In addition, the coupled differential equations can show
the complete spectrum of exact interdimensional degeneracies of
two-electron system \cite{dun4,dun5}.

Despite the achievements outlined above, the method by Dunn and
Watson \cite{dun2,dun3,dun4,dun5} seems quite complicated. To
apply the methods of dimensional scaling to
higher-angular-momentum states, a formalism needs to be developed
which factors the D-dimensional rotational degrees of freedom from
the internal degrees of freedom. They have to make several steps
to decompose an $n$-rank tensor to irreducible tensors by making
use of the method of group theory \cite{wey,boe}. They first
decomposed an $n$-rank tensor space into $(a+1)$-fold traceless
tensor subspaces. Then, they applied the Young operators on the
one-fold traceless tensor space to project it into the irreducible
invariant tensor subspace. At last, they analyze the equivalence
of those representations and pick up the independent basis of the
angular momentum belonging to the given irreducible representation
by the generalized Schwartz identities. Each step in the 
decomposition is very complicated.

It is worth pointing out that the set of variables $r^{(1)}$,
$\cdots$, $r^{(N)}$, $r^{(1, 2)}$, $r^{(1, 3)}$, $\cdots$,
$r^{(N-1, N)}$ is not a complete set of internal variables even in
three dimensions \cite{gu2} because they cannot describe the
configuration uniquely. One can convince himself by the following
counter example. For three-electron system in three-dimensions the
six variables $r^{(1)}$, $r^{(2)}$, $r^{(3)}$, $r^{(1,2)}$,
$r^{(1,3)}$, and $r^{(2, 3)}$ cannot distinguish two states
related by spatial inversion in the body-fixed frame.

Another complicated problem occurs in the formalism for N-electron
system in $D$-dimensions. Since a right-hand rectangular
coordinate frame in $D$-dimensions is determined by $(D-1)$
vectors which are linearly independent, the situations for
$N$-electron system with $N>D-1$ and $N\leq D-1$ are very
different \cite{gu2}. On the other hand, the possible irreducible
representations of SO($D$) are different for $D>2N$, $D=2N$, and
$D<2N$. Unfortunately, the complicated situations were not
discussed in the formalism \cite{dun2,dun3} in detail.

In our recent paper \cite{gu1}, we presented a new method of
separating the rotational degrees of freedom from the internal
degrees of freedom for three-body system in $D$-dimensions. After
removing the motion of the center of mass there are two Jacobi
coordinate vectors for a three-body system, just like two position
vectors in two-electron system where the mass of nucleus is
assumed to be infinite. An angular momentum state for a three-body
system in $D$ dimensions is described by the irreducible
representation of SO($D$) denoted by a two-row Young diagram
$[\mu, \nu]$. Two states in three dimensions with the same angular
momentum $l$ and the different parities belong to the same
irreducible representation of SO(3), but in $D$-dimensions where
$D>3$ they belong to different irreducible representations denoted
by the Young diagrams $[l,0]$ and $[l,1]$ of SO($D$). We
explicitly find the independent base-function for the highest
weight state of each irreducible representation $[\mu,\nu]$, which
is a homogeneous polynomial of the components of two Jacobi
coordinate vectors and satisfies the Laplace equation. Its
partners can be calculated from it by the lowering operator. Any
wave function of this system with a given angular momentum
$[\mu,\nu]$ can be expressed as a combination of the
base-functions where the coefficients are called the generalized
radial functions. The generalized radial equations satisfied by
the generalized radial functions are easily derived explicitly
\cite{gu1}. The general interdimensional degeneracies for any
angular momentum $[\mu,\nu]$ can be found from the generalized
radial equations, which is the main purpose of this Letter.

The plan of this Letter is as follows. In Sec. II, we outline the
method given in our recent work and emphasize the derivation for
the generalized radial equations for a quantum three-body system
in $D$-dimensions. The reader is suggested to refer our published
papers \cite{gu2,gu1} for the detail. The interdimensional
degeneracies for this system will be demonstrated from the
generalized radial equations in Sec. III. Some conclusions will be
given in Section IV.

\section{Generalized radial equations in $D$-dimensions}

In our recent paper \cite{gu2} we separated completely the global
rotational degrees of freedom in the Schr\"{o}dinger equation for
an $N$-body system in the three-dimensional space from the
internal ones. We have determined a complete set of $(2l+1)$
independent base-functions for a given total orbital angular
momentum $l$, which are the homogeneous polynomials in the
components of the Jacobi coordinate vectors and do not contain the
Euler angles explicitly. The generalized radial equations which
depend solely on internal variables are established. For the
typical three-body system in three dimensional space, such as a
helium atom \cite{duan1,duan2,duan3} and a positronium negative
ion \cite{duan4}, the generalized radial equations have been
solved numerically with high precision. This method was
generalized to the arbitrary dimensional space for a three-body
system.

After removing the motion of center of mass, the Schr\"{o}dinger
equation with a spherically symmetric potential $V$ in
$D$-dimensions can be expressed in terms of the Jacobi coordinate
vectors ${\bf x}$ and ${\bf y}$, where the atomic units
($e=\hbar=1$) are used for simplicity, \cite{gu1}:
$$\begin{array}{c}
\left\{\bigtriangledown^{2}_{\bf x} +\bigtriangledown^{2}_{\bf
y}\right\}\Psi({\bf x, y})
=-2\left\{E-V(\xi_{1}, \xi_{2}, \xi_{3})\right\}\Psi({\bf x, y}), \\
{\bf x}=\left[\displaystyle {m_{1}m_{2} \over
m_{1}+m_{2}}\right]^{1/2}
\left\{{\bf r}_{2}-{\bf r}_{1}\right\},\\[2mm]
{\bf y}=\left[\displaystyle {(m_{1}+m_{2})m_{3} \over
m_{1}+m_{2}+m_{3}} \right]^{1/2}\left\{{\bf r}_{3}-\displaystyle
{m_{1}{\bf r}_{1}+m_{2}{\bf r}_{2}\over m_{1}+m_{2}}\right\}, \\
\xi_{1}={\bf x\cdot x}, ~~~~~~ \xi_{2}={\bf y\cdot y},
~~~~~~\xi_{3}={\bf x\cdot y},
\end{array} \eqno (1) $$

\noindent
where ${\bf r}_{j}$ and $m_{j}$ are the position vector
and the mass of the $j$-th particle, respectively.

The orbital angular momentum in $D$-dimensions is described by the
irreducible representation of SO($D$), denoted by a two-row Young
diagram $[\mu, \nu]$. We only need to find the base-function of
angular momentum corresponding to the highest weight ${\bf M}$ of
the representation $[\mu,\nu]$. Its partners can be calculated
from it by the angular momentum operator $L_{ab}$ \cite{gu1}. It
was proved that the following harmonic polynomial $Q^{\mu
\nu}_{q}({\bf x, y})$ construct the independent and complete set
of the highest weight states of $[\mu,\nu]$:
$$\begin{array}{l}
Q^{\mu \nu}_{q}({\bf x, y})=\displaystyle
{X_{12}^{q-\nu}Y_{12}^{\mu-q} \over (q-\nu)!(\mu-q)!}
\left(X_{12}Y_{34}-Y_{12}X_{34}\right)^{\nu}, ~~~~~
0\leq \nu \leq q \leq \mu.~~~~~\\[2mm]
X_{12}=x_{1}+ix_{2}, ~~~~~~X_{34}=x_{3}+ix_{4}, ~~~~~~
Y_{12}=y_{1}+iy_{2}, ~~~~~~Y_{34}=y_{3}+iy_{4}.
\end{array} \eqno (2) $$

\noindent
 The formula for $Q^{\mu \nu}_{q}({\bf x, y})$ holds for
$D=3$ ($x_{4}=y_{4}=0$, $\nu=0$ or $1$) \cite{gu1} and $D>4$. When
$D=4$ we denote the highest weight states by $Q^{(S)\mu
\nu}_{q}({\bf x, y})$ and $Q^{(A)\mu \nu}_{q}({\bf x, y})$ for the
selfdual representations and the antiselfdual representations,
respectively:

$$\begin{array}{c}
Q^{(S)\mu \nu}_{q}({\bf x, y})=\displaystyle
{X_{12}^{q-\nu}Y_{12}^{\mu-q} \over (q-\nu)!(\mu-q)!}
\left(X_{12}Y_{34}-Y_{12}X_{34}\right)^{\nu}\\[2mm]
Q^{(A)\mu \nu}_{q}({\bf x, y})=\displaystyle
{X_{12}^{q-\nu}Y_{12}^{\mu-q} \over (q-\nu)!(\mu-q)!}
\left(X_{12}Y_{34}^{\prime}-Y_{12}X_{34}^{\prime}\right)^{\nu}\\[2mm]
X^{\prime}_{34}=x_{3}-ix_{4}, ~~~~~~Y^{\prime}_{34}=y_{3}-iy_{4}.
\end{array} \eqno (3) $$

Obviously, the base-function $Q^{\mu\nu}_{q}({\bf x, y})$ is a
homogeneous polynomial of degree $q$ and $(\mu+\nu-q)$ in the
components of the Jacobi coordinate vectors ${\bf x}$ and ${\bf
y}$, respectively, and does not contain any angle variables
explicitly. It is easy to check that $Q^{\mu\nu}_{q}({\bf x, y})$
satisfies following formulas:

$$\begin{array}{l}
\bigtriangledown_{\bf x}^{2} Q_{q}^{\mu \nu}({\bf x, y})
=\bigtriangledown_{\bf y}^{2} Q_{q}^{\mu \nu}({\bf x, y})
=\bigtriangledown_{\bf x}\cdot \bigtriangledown_{\bf y}
Q_{q}^{\mu \nu}({\bf x, y})=0, \\
{\bf x}\cdot \bigtriangledown_{\bf x} Q_{q}^{\mu \nu}({\bf x, y})
=qQ_{q}^{\mu \nu}({\bf x, y}), \\
{\bf y}\cdot \bigtriangledown_{\bf y}Q_{q}^{\mu \nu}({\bf x, y})
=(\mu+\nu-q)Q_{q}^{\mu \nu}({\bf x, y}), \\
{\bf y}\cdot \bigtriangledown_{\bf x}Q_{q}^{\mu \nu}({\bf x, y})
=(\mu-q+1)Q_{q-1}^{\mu \nu}({\bf x, y}), \\
{\bf x}\cdot \bigtriangledown_{\bf y}Q_{q}^{\mu \nu}({\bf x, y})
=(q-\nu+1)Q_{q+1}^{\mu \nu}({\bf x, y}). \end{array} \eqno (4) $$

Any wave function of this system corresponding to the highest
weight ${\bf M}$ of the representation $[\mu,\nu]$ can be
expressed as a combination of the base-functions $Q^{\mu
\nu}_{q}({\bf x, y})$,

$$\Psi^{[\mu,\nu]}_{\bf M}({\bf x,y})=\displaystyle \sum_{q=\nu}^{\mu}~
\psi^{\mu\nu}_{q}(\xi_{1}\xi_{2}\xi_{3})Q^{\mu \nu}_{q}({\bf x,
y}), \eqno (5) $$

\noindent
 where the coefficients
$\psi^{\mu\nu}_{q}(\xi_{1}\xi_{2}\xi_{3})$ are called the
generalized radial functions. When applying the Laplace operator
$\bigtriangledown_{\bf x}^{2}+\bigtriangledown_{\bf y}^{2}$ to the
wave function $\Psi^{[\mu,\nu]}_{\bf M}({\bf x,y})$, the
calculation consists of three parts. The first is to apply the
Laplace operator to the generalized radial functions
$\psi^{\mu\nu}_{q}(\xi_{1}\xi_{2}\xi_{3})$, which can be
calculated by replacement of variables. The second is to apply it
to $Q_{q}^{\mu \nu}({\bf x, y})$, which is vanishing due to Eq.
(4). The last part is the mixed application
$$\begin{array}{l}
2\left\{(\partial_{\xi_{1}}\psi^{\mu\nu}_{q}) 2{\bf x}+
(\partial_{\xi_{3}}\psi^{\mu\nu}_{q}) {\bf y}\right\}\cdot
\bigtriangledown_{\bf x}Q_{q}^{\mu \nu}
+2\left\{(\partial_{\xi_{2}}\psi^{\mu\nu}_{q}) 2{\bf y}+
(\partial_{\xi_{3}}\psi^{\mu\nu}_{q}) {\bf x}\right\}\cdot
\bigtriangledown_{\bf y}Q_{q}^{\mu \nu}, \end{array} \eqno (6) $$

\noindent
 which can be calculated by Eq. (4) easily. Hence, we
obtain the generalized radial equations, satisfied by the
$(\mu-\nu+1)$ functions $\psi^{\mu\nu}_{q}(\xi_{1}\xi_{2}\xi_{3})$

$$\begin{array}{c}
\left\{
4\xi_{1}\partial^{2}_{\xi_{1}}+4\xi_{2}\partial^{2}_{\xi_{2}}
+2(D+2q)\partial_{\xi_{1}}+2(D+2\mu+2\nu-2q)\partial_{\xi_{2}}
+\left(\xi_{1}+\xi_{2}\right)\partial^{2}_{\xi_{3}} \right.\\
\left.+4\xi_{3}\left(\partial_{\xi_{1}}+\partial_{\xi_{2}}\right)
\partial_{\xi_{3}}\right\} \psi^{\mu \nu}_{q}
+2(\mu-q) \partial_{\xi_{3}}\psi^{\mu \nu}_{q+1} +2(q-\nu)
\partial_{\xi_{3}}\psi^{\mu \nu}_{q-1}
=-2\left(E-V\right) \psi^{\mu \nu}_{q}.
\end{array} \eqno (7) $$

\section{The general interdimensional degeneracies }

The radial equations (7) were derived without any approximation.
This equations for  $\psi^{\mu \nu}_{q}$ in $D+2$ dimensions is
$$\begin{array}{c}
\left\{
4\xi_{1}\partial^{2}_{\xi_{1}}+4\xi_{2}\partial^{2}_{\xi_{2}}
+2(D+2q+2)\partial_{\xi_{1}}+2(D+2\mu+2\nu-2q+2)\partial_{\xi_{2}}
+\left(\xi_{1}+\xi_{2}\right)\partial^{2}_{\xi_{3}} \right.\\
\left.+4\xi_{3}\left(\partial_{\xi_{1}}+\partial_{\xi_{2}}\right)
\partial_{\xi_{3}}\right\} \psi^{\mu \nu}_{q}
+2(\mu-q) \partial_{\xi_{3}}\psi^{\mu \nu}_{q+1} +2(q-\nu)
\partial_{\xi_{3}}\psi^{\mu \nu}_{q-1} =-2\left(E-V\right)
\psi^{\mu \nu}_{q}.
\end{array}  $$

The radial equations for  $\psi^{\mu' ~\nu'}_{q'}$ in $D$
dimensions where $\mu'=\mu+1$, $\nu'=\nu+1$ and $q'=q+1$ is
$$\begin{array}{c}
\left\{
4\xi_{1}\partial^{2}_{\xi_{1}}+4\xi_{2}\partial^{2}_{\xi_{2}}
+2(D+2q+2)\partial_{\xi_{1}}+2(D+2\mu+2\nu-2q+2)\partial_{\xi_{2}}
+\left(\xi_{1}+\xi_{2}\right)\partial^{2}_{\xi_{3}} \right.\\
\left.+4\xi_{3}\left(\partial_{\xi_{1}}+\partial_{\xi_{2}}\right)
\partial_{\xi_{3}}\right\} \psi^{\mu'\nu'}_{q'}
+2(\mu-q) \partial_{\xi_{3}}\psi^{\mu'\nu'}_{q'+1} +2(q-\nu)
\partial_{\xi_{3}}\psi^{\mu'\nu'}_{q'-1}
=-2\left(E-V\right) \psi^{\mu'\nu'}_{q'}.
\end{array}  $$

It is evident that the radial equations (7) are invariant under
the replacement
$$ \mu,~ \nu,~ q,~ (D+2) \Longleftrightarrow (\mu+1),~(\nu+1),~
(q+1),~D , ~~~~~~\eqno (8) $$

\noindent yielding the complete spectrum of the exact
interdimensional degeneracies between the $[\mu, \nu]$ state in
$(D+2)$-dimensions and the $[\mu+1, \nu+1]$ state in
$D$-dimensions. The parities of both states on two sides of
correspondence (8) have the same value $(-1)^{\mu-\nu}$. In
comparison with the interdimensional degeneracies discussed in
literatures \cite{her,her2,her3,goo,goo2,dor,dun1,ros,tan},
$\mu=l$, $\nu=0$, and the parity is $(-1)^{l}$.

Consider a system where two electrons move around a nucleus, such
as a $D$-dimensional helium atom. It is a typical three-body
system where the first and second particles are chosen to be two
electrons and the third one is the nucleus. In the
non-relativistic quantum mechanics, the concept of spin comes from
the experimental results, not from the equation. As usual
\cite{her,loe}, we also assume that the spin of an electron in the
$D$-dimensional space is $1/2$, just as that in the real three
dimensional space, and the total wave function of two electrons is
antisymmetric in the permutation of them. Since the Jacobi
coordinate vector ${\bf x}$ changes its sign in the permutation
$R$ of two electrons and ${\bf y}$ keeps invariant, we obtain from Eq. (2)
that the permutation parity of the base-function $Q^{\mu
\nu}_{q}({\bf x, y})$ is $(-1)^q$:
$$R~Q^{\mu \nu}_{q}({\bf x, y})=(-1)^{q}~Q^{\mu \nu}_{q}({\bf x, y}).
\eqno (9) $$

\noindent
In the permutation of two electrons $Q^{\mu \nu}_{q}({\bf x, y})$
and $Q^{(\mu+1)(\nu+1)}_{q+1}({\bf x, y})$ have the opposite
permutation parities. Therefore, two states in the
interdimensional degeneracy (8) must have the different spin $S=1$
and $S=0$, respectively. In summary, we obtain the complete spectrum 
of the exact interdimensional degeneracies as follows. For a given 
parameter $a=0$ or $1$, all the states with the orbital angular
momentum $[\mu+n,\nu+n]$, the spin $S=[1+(-1)^{a+n}]/2$, and 
the parity $(-1)^{\mu-\nu}$ in the dimension $D-n$ are 
interdimensional degenerate where $n$ is an arbitrary integer
satisfying $D\geq n \geq -\nu$. 
These general interdimensional degeneracies include all those 
degeneracies discovered in literatures
\cite{her,her2,her3,goo,goo2,dor,dun1,ros,tan}.

\section{Conclusions }

In this Letter we have provided a systematic procedure for 
analysis of observed degeneracies among different states in 
different dimensions and yielded considerable insight into 
the energy spectra of three body system. Since the generalized
radial equations (7) for a quantum three-body system with a 
spherically symmetric potential $V$ in $D$-dimensions are derived 
without any approximation \cite{gu1}, the interdimensional
degeneracies given at the end of the preceding section are exact 
and general. This general interdimensional degeneracies for 
a three-body systems should be generalized to $N$-body system, 
which is our future task. 

Before ending this Letter we would like to make a remark.
The energy $E$ is completely determined by the solution
of the coupled differential equations (7) with a suitable
boundary conditions, although we could not solve Eq. (7)
analytically but only numerically. Since the equations 
for two states in the general interdimensional degeneracies 
are exactly the same, their energies must be the same. However, 
in the practical numerical calculations \cite{duan1,duan2,duan3,duan4}
we expanded the solutions as a Taylor series in the orthogonal 
bases, which depend on the integral in the configuration space.
As pointed out in Eq. (44) of our recent paper \cite{gu1}
the volume element of the configuration space contains a 
factor $(\xi_{1}\xi_{2}-\xi_{3}^{2})^{(D-3)/2}$, which depends
on the dimension $D$. On the other hand, a straightforward
calculation shows that the integrals over the angular variables
for the base-function $Q^{(\mu+1)(\nu+1)}_{q+1}({\bf x, y})$ 
in $D$-dimensions contains one more factor 
$(\xi_{1}\xi_{2}-\xi_{3}^{2})$ than that for the base-function 
$Q^{\mu \nu}_{q}({\bf x, y})$ in $(D+2)$-dimensions so that
two wave functions of the states in the general interdimensional 
degeneracies  have the same integral in the configuration space.


\noindent

{\bf ACKNOWLEDGMENTS}. This work was supported by the National
Natural Science Foundation of China.

\end{document}